# Optimizing Service Function Chain Mapping in Network Function Virtualization through Simultaneous NF Decomposition and VNF Placement

A.Asgharian-Sardroud[1], M. H. Izzanlou[2], A. Jabbari[3], and S. Mahmoodian Hamedani [4]

[1,2,4] Department of Electrical and Computer Engineering, Urmia University, Urmia, Iran

[3] Department of Electrical and Computer Engineering, Razi University, Kermanshah, Iran

## Abstract

Network function virtualization enables network operators to implement new services through a process called service function chain mapping. The concept of Service Function Chain (SFC) is introduced to provide complex services, which is an ordered set of Network Functions (NF). The network functions of an SFC can be decomposed in several ways into some Virtual Network Functions (VNF). Additionally, the decomposed NFs can be placed (mapped) as VNFs on different machines on the underlying physical infrastructure. Selecting good decompositions and good placements among the possible options greatly affects both costs and service quality metrics. Previous research has addressed NF decomposition and VNF placement as separate problems. However, in this paper, we address both NF decomposition and VNF placement simultaneously as a single problem. Since finding an optimal solution is NP-hard, we have employed heuristic algorithms to solve the problem. Specifically, we have introduced a multiobjective decomposition and mapping VNFs (MODMVNF) method based on the non-dominated sorting genetic multi-objective algorithm (NSGAII) to solve the problem. The goal is to find near-optimal decomposition and mapping on the physical network at the same time to minimize the mapping cost and communication latency of SFC. The comparison of the results of the proposed method with the results obtained by solving ILP formulation of the problem as well as the results obtained from the multi-objective particle swarm algorithm shows the efficiency and effectiveness of the proposed method in terms of cost and communication latency.

## 1. Introduction

In recent years, the ever-increasing proliferation of applications on computers and mobile devices has increased the demand for network services. Traditionally, network services were delivered through expensive hardware equipment that could neither keep up with the ever-increasing demand nor allow new services to be deployed cost-effectively. Network Functions Virtualization (NFV) has recently been proposed to overcome such issues. NFV is a promising network architecture concept that aims to virtualize network service by implementing single service components in virtual machines enabled on commercial servers (COTS). Each service is represented by a service function chain (SFC), which is a collection of NFs that execute according to a specific order. Each NF in SFC can be decomposed into several simpler NFs. Also, the decomposed NFs can be mapped on various physical network nodes. The selection of the best decomposition among different types of decompositions (SFC decomposition problem) and finding VNF mappings (VNF placement problem) have a great impact on the resource consumption and service quality of an SFC which means the problems of SFC decomposition and VNF placement are multi-objective problems. Therefore, it is necessary to solve these problems by using multi-objective algorithms. Heuristic and meta-heuristic algorithms are methods for solving single-objective and multi-objective problems. Among the most famous metaheuristic algorithms to solve multi-objective problems, is the non-dominant sorting genetic algorithm (NSGA-II) [1], which outperforms other multi-objective elitist algorithms (MOEA). The most important motivation for creating the NSGA-II algorithm was the high computational complexity of GA and the lack of elite selection. NSGA-II is also notable for finding a wider range of solutions for multi-objective optimization problems in a range close to the Pareto optimal set. In addition, one of the prominent features of NSGA-II is that, unlike previous approaches, it does not require any user-specific parameters to capture diversity among population members [1].

---

[1] a.asgharian@urmia.ac.ir
[2] m.h.izzanlou1996@gmail.com
[3] amin.jabbbari@gmail.com
[4] sepehrmahmoodian@gmail.com

To fulfill an SFC, both the NF decomposition and VNF placement problems should be solved. Previous works have considered these problems as separate problems and give some solutions to cope with these problems. In this approach, first, the NFs are decomposed and then they are mapped on VNFs. However, by first fixing the decomposition and then finding a placement we may miss some good solutions. In this paper, we consider fulfilling an SFC request as a single problem. As a result, we introduce an algorithm that finds decompositions and placements simultaneously.

## 1.2. SFC structure

One of the main challenges of NFV orchestration is the deployment and appropriate prototyping of sequences of VNFs that form the Service Function Chain (SFC) [2, 3]. SFC is a mechanism that allows different service functions to be connected together to form a service that enables carriers to benefit from a virtualized software-defined infrastructure. SFC is an enabler for NFV that provides a flexible and cost-effective alternative to today's static environment for Cloud Service Providers (CSP), Application Service Providers (ASP), and Internet Service Providers (ISPs). The required services usually are represented as Service Graphs (SG) in which the number, type, and order of each NF are determined according to the performance and behavioral characteristics of the corresponding service. Service decomposition is the process of converting an SG containing high-level NFs (composite) into SGs with more basic NFs that can finally be mapped onto the physical infrastructure. Service decomposition aims to reuse basic building blocks (NF), build new and more complex services, and request high-level services without worrying about detailed implementations. Primitive NFs can be of different types, which means they can be implemented through different techniques. These techniques include 1) virtual machine (VM) images with different virtualization techniques such as Xen, Virtual Box, and VMware; 2) Processing inside a container (Process); 3) package I/O drivers such as Intel's Data Plane Development Kit (DPDK), which includes a set of open source libraries integrated with VMware to speed up the processing of packages that use services such as VNF; 4) Hardware equipment. Figure 1 shows an example of a service decomposition. As seen in Figure 1, high-level NF2 is broken down into NF4 and NF5. Each of them can be further decomposed into more primitive NFs. This decomposition is determined during the design of NFs or SGs. They can be stored as a tree-like data structure in a database to be used by the mapping algorithm. The leaves of the parse tree should contain NFs that are available for low-level implementation and resource requirements. Technically, connecting NFs of the same type (for example, similar processes in containers) is less complicated than connecting non-homogeneous NFs (for example, an NF on a container and an NF based on DPDK). In addition, the higher the number of NFs of the same type, the more NFs may be placed in the same physical node. If NFs of the same type are directly connected, it leads to less resource consumption. Therefore, prioritizing service decompositions with NFs of the same type that is more directly related improves embedded performance in terms of resource consumption and enables simpler NF connections.

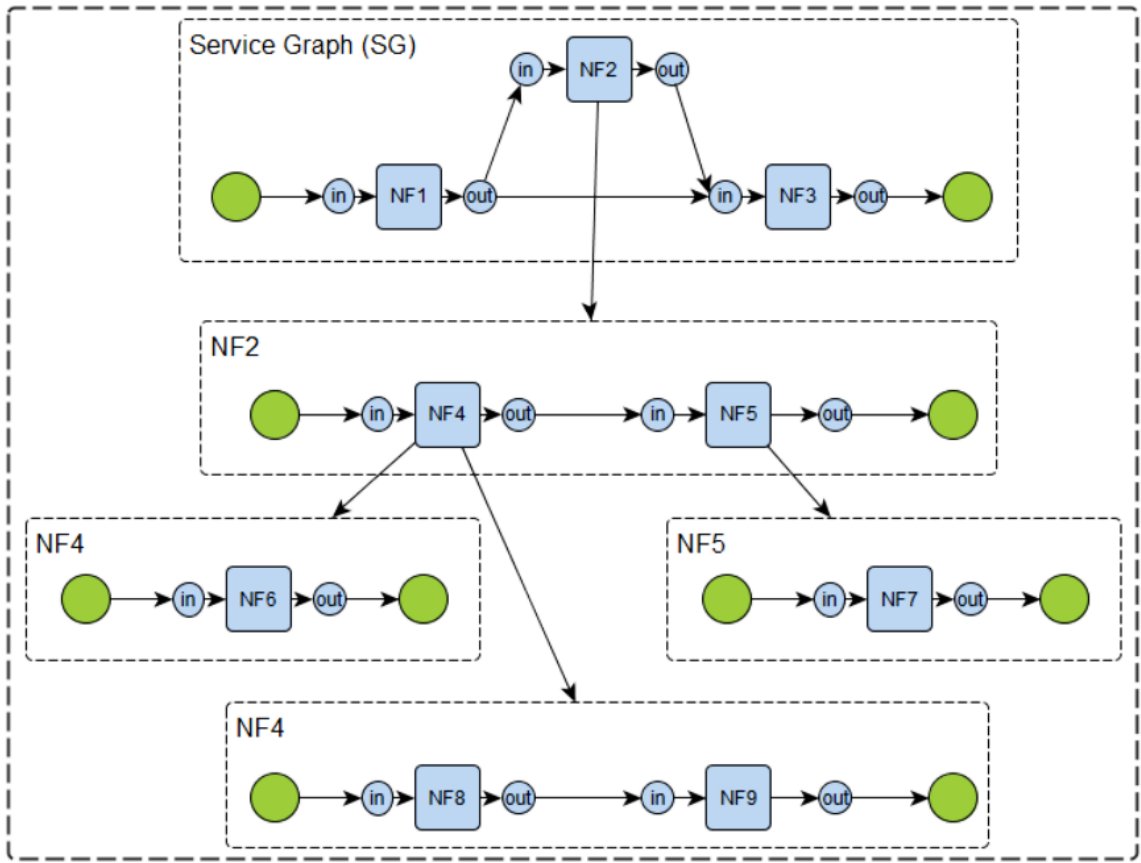

**Figure 1: An example of service decomposition [4]**

## 1.3. VNF Mapping

In NFV networks, network functions are separated from the main hardware and run as VNFs. Due to the software feature of VNFs, they can be deployed flexibly. Therefore, an important issue appears, that is, how to determine the position to place VNFs in such a way that the requirements and the quality of service are met. Such a problem is called the VNF mapping problem (VNF-P). Figure 2 illustrates the general idea of the VNFs mapping problem.

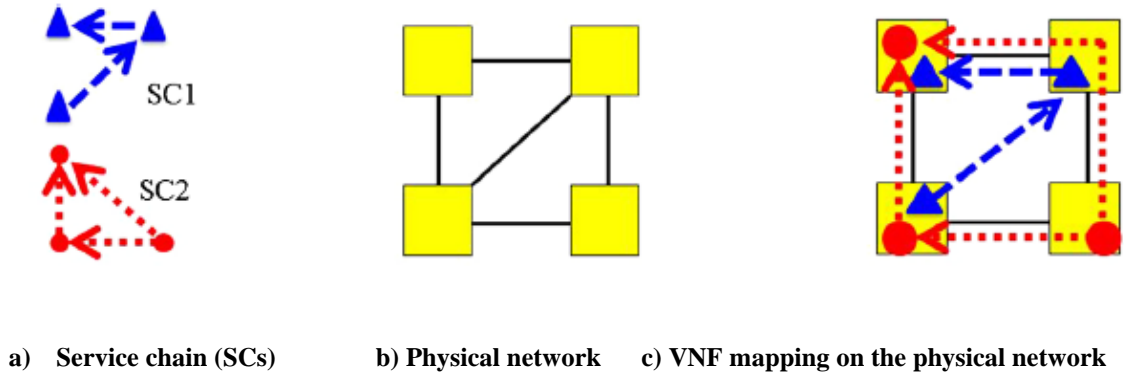

a) Service chain (SCs)       b) Physical network     c) VNF mapping on the physical network

**Figure 2: The problem of mapping VNFs on physical hardware with two service chains [1]**

In the problem of mapping VNFs, the goal is to find the optimal path for mapping SFC requests in available network resources to provide users with the requested service quality. VNF placement costs should be reduced as much as possible. Such improvements provide benefits to the network operator and promote the use of NFV technology [5]. Because the mapping problem of VNFs is an NP-hard problem [1], it is usually difficult to find the optimal solution for VNF-P, especially in large-scale network scenarios. To obtain the optimal solution for VNF-P, programming methods such as integer linear programming (ILP) and hybrid ILP (MILP) are usually used. However, due to the time complexity of VNF-P, the corresponding optimization problem cannot be solved using conventional mathematical approaches in practice [1]. Therefore, to overcome this complexity and NP-hardness, several heuristic and meta-heuristic approaches have been introduced that have remarkably short execution times, making them good candidates for real-world network optimization. In this paper, we use the MODMVNF algorithm to solve the problem of mapping VNFs. The key elements in NSGA-II are:

1) Generation of the initial population based on the comparison and limitations of the problem

2) Evaluation of the generated population according to the objective functions

3) Non-dominated sorting approach

4) Calculation of the control parameter called crowding distance

5) Selection of parental population for reproduction

6) Perform crossover and mutation operator

## 1.4. Contributions

Our proposed method simultaneously finds a near-optimal decomposition and mapping considering both cost and delay criteria, for each SG. Some research has been done on finding the best decomposition and mapping for SGs. This research usually considered only finding decomposition or mapping with a single criterion such as mapping cost or delay. Considering that the SG decomposition and placement problems have a multi-objective nature, and their optimal solutions are related to each other, we have introduced a method that finds a solution for both of these problems at the same time considering two major criteria: cost and delay.

## 2. Related works

So far, many works have been done in the area of creating service function chains and mapping these chains on physical infrastructures that we survey each one separately. In the following two subsections, we review some related works on service function decomposition and VNF mapping.

### 2.1. Creating service function chains

The VNF chaining problem (VNF-C), which is also called the service function chaining problem, mainly focuses on the chaining mechanisms of VNFs and routing the relevant traffic through these VNFs in order, to reach the destinations [6]. Intending to solve this problem, the IETF has specifically created a working group to document new approaches to service delivery and VNF functionality and SFC architecture and traffic command algorithms [7], which prove the importance of SFC. To solve this problem, Sahaf et al. [4], and Li et al. [8] formulated it as an ILP model with different objectives. Sahaf et al. aimed to minimize the total cost by judicious choice of VNF decomposition, but Lee et al. aimed to minimize end-to-end delay by random flow distribution. Sahaf et al. formulated the VNF-C problem as an ILP model, which was solved by COIN-OR solver [9]. Their algorithm can be divided into two steps, namely, decomposition selection and mapping. In the decomposition selection step, a decomposition with the minimum cost is selected and in the mapping step, it is placed on an appropriate node. Li et al. also first proposed two criteria: uniform distribution scheme (flows are distributed equally to all VNF instances and no resource constraints are considered) and network distribution scheme (flows are distributed based on the delay between VNF samples). Then, by changing the flow distribution, it was claimed that the two measures reduced the overall delay by 27.14-40.56% and 12.77-28.84%, respectively.

For practical situations, solving the VNF-C problem dynamically is more suitable. In this regard, Liu et al. [10] considered the problem of re-adapting VNF-C in a dynamic environment so that, on the one hand, it should satisfy the new service requirements. On the other hand, it should readjust the existing service chain to meet the changing needs of users. To solve these two cases, Liu et al. first formulated the corresponding ILP model that can be used to obtain the optimal solution. However, due to the huge execution time spent to solve the ILP model, they also proposed an innovative method based on the idea of column generation (CG) that only needs to generate variables that may improve the objective formulated in the ILP model. In addition, the main idea behind the CG-based algorithm was to decompose the original problem into a main problem and a sub-problem and then solve them iteratively to obtain the near-optimal solution. For flexibility and economic reasons, many service providers want to consider VNF-C and VNF placement. To achieve such a common goal [11], Botton and colleagues proposed a model for VNF-C that considered not only resource capacity constraints but also VNF location constraints. Similarly, the corresponding proposed heuristic algorithm was also implemented under such constraints. However, it was found that this proposed innovative method considered the VNF-C problem as a VNE problem and therefore the problem actually solved was VNE instead of VNF-C. Although the VNF-C problem is similar to the VNE problem, for VNF-C, service request topology may change according to the deployment of VNFs. Most of the works mentioned above studied the VNF-C problem in the context of SDN, that is, they usually solved the VNF-C problem in a centralized manner. However, distributed solutions for VNF-C are also interesting in some applications. Doro et al. [12] implemented a distributed service chaining algorithm. The solution proposed by Duro et al. was based on the theory of non-cooperative games, which showed that there is a game with competition between individual players and only self-sufficient alliances are possible in the game. Furthermore, for the selfish and competitive behavior of customers, the service chain composition was formulated as an atomic-weight congestion game that processed the weighted potential function and adopted a Nash equilibrium (a state where no player can profit just by changing strategies). From the practical view, service requirements may change over time. For example, VNFs may be added or removed from the service chain according to dynamic needs. As such, methods of recombining services, remapping functions, and rescheduling are essentially required. In [13], a method to solve the scaling problem of VNF-C has been proposed, to solve such a case, two heuristic methods were proposed. The first method adds or removes VNFs based on the reserved function path, and the second method aims to optimize the reserved function path according to the current state of the network. Despite the two proposed heuristic methods, their joint use can have better results than using each of them exclusively. Lee et al. [14] considered the problem of creating a chain and choosing a path of VNF services as a Grey theory problem. Based on the concept of systematization theory, the degree of relationality is used to measure the relationship between the candidate service composition, the ideal service composition, and the worst service combination. Then, a membership function was used to calculate the membership of the candidate composition in terms of the ideal service position. Among all these solutions, finally, the best solution is chosen. Although they claimed to achieve 200ms average latency and 8% lower packet drop than the random method, each time a request is chosen to reach the best solution, it computes all possible paths that lead to a huge computational burden. Wang et al. [15] converted the VNF-C problem into a Markov chain model. Then, based on the existing approximate Markov method [16], they proposed another distributed algorithm to achieve optimal solutions. Nam et al. [17] transformed the VNF-C problem into a VNF clustering and assignment problem. First, they classified VNFs according to their popularity and then assigned them to service requests according to their needs. The VNF-C problem has also been studied in many other cases such as cloud computing [18], data centers [19], carrier-grade networks [20], etc.

## 2.2. VNFs mapping

VNF mapping is a step for the optimal allocation of VNFs in the network infrastructure. Moen and Mones [21] presented a formal VNF mapping problem for resource allocation in hybrid network environments where network functions can be allocated over physical hardware and virtual instances. With the advancement of 5G networks, low and predictable end-to-end latency is becoming increasingly important as a critical factor for many applications. They have presented a general model for the efficient mapping of virtual network functions. The simultaneous mapping of VNFs is used to form a service function chain (SFC) and a chain of VNFs, and then admission control (AC) is used to reach the maximum performance state. In fact, the main problem of [21] is to present a system model that formulates the problem of allocating the desired resources for different types of SFCs and deals with the computational complexity of the problem. In [22] Bari et al. introduced the VNF orchestration problem, which was equivalent to VNF-P, and formulated it as an ILP model, in terms of OPEX minimization and network utilization maximization. Riggio et al. formulated VNF-P [23] in a radio access network scenario as an ILP model and solved it to obtain the optimal solution of VNF mapping under radio resource constraints. This problem also has been considered for different objectives. For example, Luizelli et al. [24] formulated the ILP model to minimize the end-to-end delay and resource overage ratio, while Gupta et al. [25] aimed to minimize the bandwidth consumption using the ILP model. Although they have different aims and work in different scenarios, the constraints of VNF-P, which include resource allocation, VNF mapping, and traffic engineering (TE), are generally the same.

However, the ILP model is only suitable for the condition that all variables are integers. Therefore, for some special circumstances, MILP is used instead. Addis et al. [26] proposed a VNF-P model that considers both minimizing the number of CPUs used by creating VNFs and minimizing the risk of sudden bottlenecks in network links (TEs). But, to achieve these two goals at the same time, some non-integer variables must be introduced. Therefore, the VNF-P model proposed by them was a MILP model that described the relationship between VNF mapping and traditional routing. By solving this model, Addis et all. claim to have achieved a 70% NFVI cost savings and a 5% increase in link utilization compared to studies that only consider the TE target. ILP and MILP models are usually solved by open-source optimization software such as CPLEX [27], LINGO [28], and GLPK [29] to obtain the optimal solution. Among other classical algorithms that solve two mathematical models are: branch and bound, branch and cut, etc. However, these mathematical proposals suffer from the weakness of scalability, that is, they cannot be implemented in large-scale networks, because their execution time grows exponentially with the size of the network. For example, Bari et al. [22] spent 1595.12 seconds solving the ILP model using CPLEX for a network containing only 23 nodes and 43 links. They also proposed a heuristic algorithm that solved the problem on the same network in 0.442 seconds. Due to the fact that the execution time of the heuristic method is much less than the optimal solvers and they can find solutions close to the optimal solutions, they are more common in solving VNF decomposition and mapping problems. In [23], Riggio et al. proposed a three-step heuristic algorithm for VNF-P in wireless networks. The steps of their algorithm were: calculating the candidate substrate nodes for virtual nodes, sorting the virtual nodes, and mapping them on the substrate nodes. After three stages, they claimed to have achieved an approximation of the desired performance. This turns VNF-P into a VNE problem. However, despite the similarities between VNF-P and VNE, they are different problems. Interested readers can refer to [30] to study more on this topic.

## 3. The Problem description

A service consists of several service functions that are traditionally implemented by intermediate nodes. The service is described by a chain of high-level network functions (NF) and predefined parameters called service graphs (SGs). A service graph is a set of NVFs that must be mapped onto the physical infrastructure.

Let *SG* denote the set of all service graphs that have already been sent to an Internet Service Provider (ISP). Each $sg_k \in SG$ is an independent service graph that contains several different VNFs and each of them is associated with resource requirements such as processing capacity (*C*), memory capacity (*M*), and storage capacity (*S*). An $sg \in SG$ may have multiple decompositions which we denote them $Decomp_{sg}$. That is:

$$\forall \, sg_k \in SG: Decomp_{sg_k} = \{dc_1, dc_2, \dots, dc_n\}.$$

Each decomposition is represented as a graph showing the dependency between different NFs. Therefore, NFs in decomposition are represented as nodes connected through links in the graph. We assume that the edges of the graph have a very large bandwidth capacity and do not impose any constraints on the VNF-P problem. In other words, our focus is more on node constraints rather than link constraints. Each NF in the decomposition graph $G_{dc} = (N_{dc}, L_{dc})$ has some requirements in terms of processing (*c*), memory (*m*), and storage (*s*). These requirements for each NF in the decomposition are defined as follows:

$$\forall i \in N_{dc}: \quad c_i, m_i, s_i \in N^+.$$

We use the name of used implementation techniques as NF types. If NF is implemented using the virtual machine image technique, it is of VM type. If it is implemented using the processing technique inside a container, it is of process type. If it is implemented using I/O package drivers, it is from I/O type and if implemented using hardware equipment, we name NF from the hardware type. That is:

$$\forall i \in N_{dc}: \quad t_i \in \{VM, process, I/O, \text{hardware}\}.$$

### 3.1. Physical network

The physical infrastructure of the problem is considered an undirected graph $G_p$ with nodes $N_p$ and links $L_p$. That is:

$$G_p = (N_p, L_p).$$

We use $C_{u_j}, M_{u_j}$, and $S_{u_j}$ respectively to denote the processing capacity, memory capacity, and storage capacity of the node $u_j$:

$$\forall u_j \in N_p: \quad C_{u_j}, M_{u_j}, S_{u_j} \in N^+.$$

$$\forall u_j \in N_p: \quad C_{Cost_{u_j}}, M_{Cost_{u_j}}, S_{Cost_{u_j}} \in N^+.$$

We assume that each physical node can host different types of NFs including a) virtual machine (VM) images; b) processing in a container; c) packet I/O drivers and d) hardware appliances.

### 3.2. MILP formulation of the problem

In this section, we give MILP formulation of the problem which contains the definition of decision variables, objective functions, and constraints of the problem. Table 1 shows the symbols, variables, and parameters used in our formulation.

**Table 1: Symbols, variables, and parameters used**

| | |
|---|---|
| $SG = \{sg_1, sg_2, ...\}$ | The set of all SGs that have already been sent to the ISP |
| $N = \{u_1, u_2, ...\}$ | The set of physical nodes |
| $L(i, j)$ | Path between two physical nodes *i* and *j* |
| $l(i,j)$ | Link between two physical nodes $i$ and $j$ |
| $u_i$ | A given physical host |
| $N_v$ | Total number of service chain functions |
| $N_H$ | Total number of physical hosts |
| $c_i$ | Computation usage of the $i^{th}$ NF |
| $m_i$ | Memory usage of the $i^{th}$ NF |
| $s_i$ | Storage usage of the $i^{th}$ NF |
| $C_i$ | Computation capacity of the physical node |
| $M_i$ | Memory capacity of the physical node |
| $S_i$ | Storage capacity of the physical node |
| $x_{u_j}^i$ | A binary decision variable which equals 1 if $i^{th}$ NF is mapped on physical node $u_i$ and is zero otherwise |
| $u_i$ | A binary decision variable which equals 1 if the physical node $u_j$ is selected and is zero otherwise |
| $z^{dc}$ | A binary decision variable which equals 1 if the decomposition dc is selected and is zero otherwise |
| $C_{u_i}^{free}$ | Available computation capacity of the physical node $u_i$ |
| $M_{u_i}^{free}$ | Available memory capacity of the physical node $u_i$ |
| $S_{u_i}^{free}$ | Available storage capacity of the physical node $u_i$ |
| $n$ | Population size |
| $P_t$ | The set of parents' population |
| $Q_t$ | The set of offspring population |

### 3.2.1. Decision variables

The decision variables are as:

i) The decision variable $x_{u_j}^i$, which takes the value of one if the i-th NF is written on the physical node $u_j$, and the value of zero otherwise:

$$x_{u_j}^i \in \{0,1\} \quad \forall dc \in Decomp_{SG}, \forall u_j \in N_p, \forall i \in N_{dc}.$$

ii) The decision variable $u_j$ which takes the value of one if the physical node is selected and takes the value of zero otherwise:

$$u_j \in \{0,1\}, \quad \forall u_j \in N_p.$$

iii) The decision variable $z^{dc}$, which takes the value of one if the dc decomposition is chosen, and zero otherwise.

$$z^{dc} \in \{0,1\}, \quad \forall dc \in Decomp_{SG}.$$

### 3.2.2. Objective functions

According to the decision variables, the objective functions for a unique SG can be formulated as follows:

Minimize:

$$f_1 = \sum_{u_j \in N_{p(VM)}} \sum_{i \in N_{dc}(VM)} cost(i, u_j)$$

$$+ \sum_{u_j \in N_{p(PRC)}} \sum_{i \in N_{dc}(PRC)} cost(i, u_j)$$

$$+ \sum_{u_j \in N_{p(I/O)}} \sum_{i \in N_{dc}(I/O)} cost(i, u_j)$$

$$+ \sum_{u_j \in N_{p(HW)}} \sum_{i \in N_{dc}(HW)} cost(i, u_j), \quad (1)$$

Minimize:

$$f_2 = \sum_{u_j \in N} \sum_{u_m \in N} \sum_{i \in N_{dc(type)}} L(j,m) x_{u_j}^i x_{u_m}^{i+1}, \quad (2)$$

where

$$cost(i, u_j) = \left(c_i \times C_{cost_{u_j}} + m_i \times M_{Cost_{u_j}} + s_i \times S_{Cost_{u_j}}\right) \times x_{u_j}^i.$$

In the objective function $f_1$, the total costs of using processing resources, RAM, and storage memory are calculated by the decomposition NFs selected from SG, and the best decomposition with the lowest cost is selected and mapped, and in the objective function $f_2$, the delay between used physical nodes by SG is minimized.

### 3.2.3. Constraints

In order to guarantee that the physical nodes have the capacity to accept NF, we consider some restrictions.

**i) Constraints of physical nodes**

For each physical node, the sum of requirements of all mapped NFs should not exceed the capacity of the node. Therefore, we must consider a constraint for each of the processing, memory, and storage capacities.

$$\sum_{i \in N_{dc(type)}} c_i x^i_{u_j} \leq C_{u_j}, \qquad (3)$$

$$\forall dc \in Decomp_{SG}, \forall i \in N_{dc(type)}: \quad \forall type \in \{VM, PRC, I/O, HW\}, \forall u_j \in N_p,$$

$$\sum_{i \in N_{dc(type)}} m_i x^i_{u_j} \leq M_{u_j}, \qquad (4)$$

$$\forall dc \in Decomp_{SG}, \forall i \in N_{dc(type)}: \quad \forall type \in \{VM, PRC, I/O, HW\}, \forall u_j \in N_p,$$

$$\sum_{i \in N_{dc(type)}} s_i x^i_{u_j} \leq S_{u_j}, \qquad (5)$$

$$\forall dc \in Decomp_{SG}, \forall i \in N_{dc(type)}: \quad \forall type \in \{VM, PRC, I/O, HW\}, \forall u_j \in N_p,$$

**ii) Decomposition constraint**

$$\sum_{u_j \in N_{p(type)}} x^i_{u_j} = z^{dc}, \qquad (6)$$

where

$$\forall dc \in Decomp_{SG}, \forall i \in N_{dc(type)}: \quad \forall type \in \{VM, PRC, I/O, HW\}$$

This constraint ensures that all selected decomposition NFs are mapped only once.

**iii) Constraint of delay between physical nodes**

$$\sum_{u_j \in H} \sum_{u_m \in H} \sum_{i \in N_{dc(type)}} l(j,m) x^i_{u_j} x^{i+1}_{u_m} \leq L_{target} \qquad (7)$$

## 4. The Proposed algorithm

In this section we introduce our multiobjective algorithm for simultanous decomposition selection and VNFs mapping based on well-known NSGAII algorithm.

Our algorithm solves both the NF decomposition and VNF placement problems simultaneously. The steps of the algorithm are described in detail below.

### 4.1. Gene, chromosome

First, we determine the gene and chromosome structure. The structure of a typical gene is shown in Table 2. Each gene contains an NF identifier and a physical node identifier. In other words, each gene indicates which NF is mapped to which physical node. We denote the total number of NFs in the system by $N_v \times SG$. Therefore, each chromosome is an array of size $N_v \times SG$ that contains genes. Each individual chromosome is a feature of a larger structure, the "solution" which is depicted in Figure 3.

Table 2: Structure of class "gene"

| Name | Type of data |
|---|---|
| Node_ID | Integer |
| NF_ID | Integer |
| Dec_ID | Integer |
| SG_ID | Integer |

### 4.2. Initial and offspring population

Let $P_t$ and $Q_t$ be the population of parents and offspring, respectively. To build $P_t$, first, according to the decompositions related to an *SG*, we make a two-dimensional matrix with the number of rows equal to the number of decompositions and the number of columns equal to the maximum number of VNFs in the decompositions. Then, one of the rows of the mentioned matrix is randomly selected, which indicates the corresponding decomposition. According to the selected row, computation resources and RAM memory, and storage memory based on the type of VNF participating in the decomposition, are allocated to that VNF as required requirements. Then, the determined VNFs are mapped on the physical nodes of the network, which will be an acceptable solution to the considered problem.

Now, by creating *N* samples of these solutions, the set $P_t$ is built, and the set $Q_t$ is created using crossover and a binary tournament choice strategy.

Figure 3 shows the structure of $P_t$ and $Q_t$ :

$$P_t = \{sol_1, sol_2, \ldots, sol_N\}$$

and

$$Q_t = \{offspring_1, offspring_2, \ldots, offspring_N\}$$

### 3.3.3. Sorting the population and calculating the crowding distance

We use the non-dominated sorting algorithm [31] to rank the input population. After sorting the population, the crowding distance parameter is calculated for each member of the population, and the population in terms of rank and crowded distance is sorted. Figures 4 and 5 show the process of the non-dominated sorting algorithm and calculating the crowding distance.

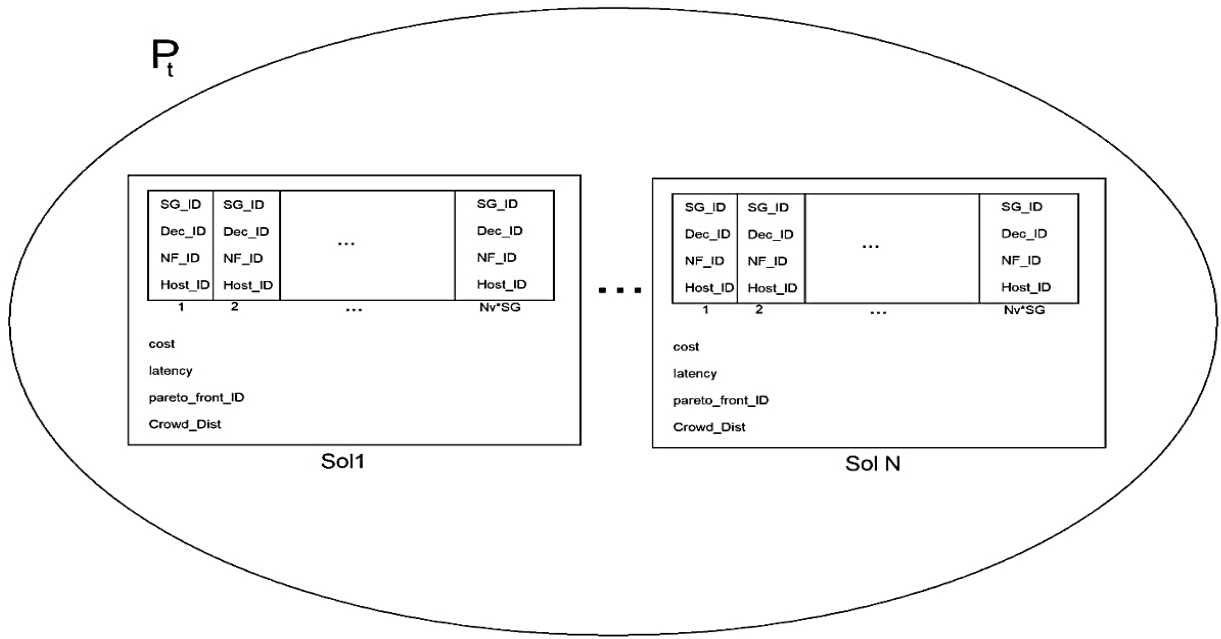

**Figure 3: The structure of a population sample**

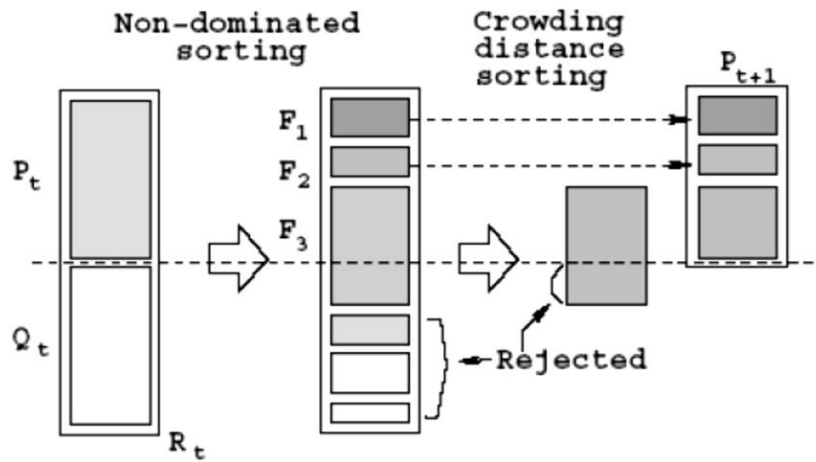

**Figure 4: The non-dominated algorithm (Deb ae al. [31])**

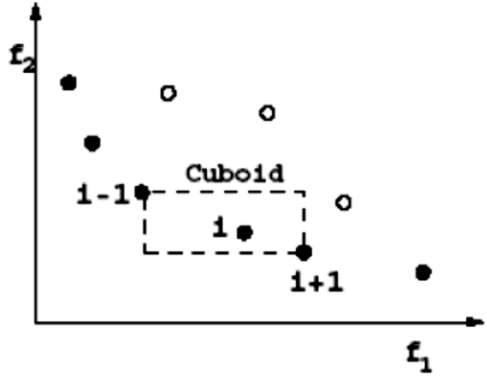

**Figure 5: The crowding distance (Deb et al. [31])**

### 3.3.4. Crossover, mutation, and fitness

The binary tournament selection method is used to randomly select parents from the total $P_t$. In this way, two members $sol_1$ and $sol_2$ are selected from the set $P_t$. If the two selected parents have the same decomposition, the crossover will be done on them, otherwise, these two parents will have no offspring. In the crossover process, each parent chromosome is divided into four equal-sized segments. Then, for each segment, a random selection is made between the two parents. In this way, the offspring's chromosomes are obtained by random shuffling of their parent chromosomes. Finally, for each offspring, the function calc_fitness is called to calculate its fitness value. The pseudocode of the calc_fitness function is shown in Algorithm 1. Now, each offspring must be passed as a parameter to the mutation function. The pseudocode of this algorithm is also shown in Algorithm 2. This function takes a solution value and checks for each gene whether the suggested resources in the gene are suitable for VNF mapping. If the allocation is practical, the VNF resource requirement is subtracted from the remaining resources of the designated node. If all the genes of the chromosome show functional assignments, the objective functions $f_1$ and $f_2$ introduced in equations (1) and (2), are calculated. Otherwise, the fitness values are set to infinity. The utility level for the minimization functions $f_1$ and $f_2$ is set to the large number $+\infty$. This approach prevents impossible solutions from prevailing over other solutions. The routine in which the above-mentioned functions are called is the make_new_pop function, whose pseudocode is included in Algorithm 3. This function receives the main population $P_t$ in round t as input and selects two solutions $sol_1$ and $sol_2$ from the set $P_t$. Then, by calling the crossover function, two children $offspring_1$ and $offspring_2$ are created and the mutation operator is applied to them, and then these two offspring are added to the population of $Q_t$ children. Finally, the $Q_t$ array is returned by the make_new_pop function (Algorithm 3).

**Algorithm 1: Pseudo-code of function calc_fitness**

*Input*: A solution sol

*Output*: Fitness of the solutions sol if it is feasible and infinity otherwise

1:    feasible =true;
2:    for index=1 to $N_v$ do
3:       $u_i = sol.chromosome[index].host\_Id$;
4:       $i = sol.chromosome[index].VNF\_Id$;
5:       if $(c_i \leq C_{u_i}^{free})$ & $(m_i \leq M_{u_i}^{free}))$ & $(s_i \leq S_{u_i}^{free})$ then
6:          $C_{u_i}^{free} = C_{u_i}^{free} - c_i$;   $M_{u_i}^{free} = M_{u_i}^{free} - m_i$;   $S_{u_i}^{free} = S_{u_i}^{free} - s_i$;
7:          sol.latency= sol.cost $=+\infty$ ;
8:       else
9:          feasible=false;
10:         break;
11:     end if
12:   end for
13:  if (feasible==true) then
14:  sol.latency= 0;
15:  sol.cost=0;
16:  for each $sg_k \in SG$ do
17:     for each $u_i \in H$ do
18:         for each $i \in N_{dc}$ do
19:         
$$sol.cost = sol.cost + \sum_{u_j \in N_{p(VM)}} \sum_{i \in N_{dc}(VM)} cost(i, u_j) + \sum_{u_j \in N_{p(PRC)}} \sum_{i \in N_{dc}(PRC)} cost(i, u_j) + \sum_{u_j \in N_{p(I/O)}} \sum_{i \in N_{dc}(I/O)} cost(i, u_j) + \sum_{u_j \in N_{p(HW)}} \sum_{i \in N_{dc}(HW)} cost(i, u_j) \ ;$$

20:         
$$sol.latency = \sum_{i \in N_{dc(type)}} \sum_{u_i \in N} \sum_{u_m \in N} L(i,m) x_{u_j}^{i} x_{u_m}^{i+1} \ ;$$

21:         end for
22:     end for
23:  end for
24: else

| 25: | sol.latency=+∞; |
| 26: | sol.cost =+∞; |
| 27: | end if |
| 28: | return sol |

**Algorithm 2: Pseudo-code of procedure mutation**

*Input: an offspring*

*Output: a mutated offspring*

| 1: | Iterations = offspring.num_used_hosts; |
| 2: | for count = 1 to iterations/3 do |
| 3: | j = random (1 ... $N_v$); |
| 4: | k = random (1 ... $N_v$); |
| 5: | swap (offspring.chromosome.[j].host_ID , offspring.chromosome.[k].host_ID) |
| 6: | end for |
| 7: | calc_fitness (offspring); |
| 8: | if offspring.latency =+∞ then |
| 9: | go to line 2 |
| 10: | end if |

**Algorithm 3: Pseudo-code of procedure make_new_pop**

| *Input*: $P_t$ | // parents population $P_t$ at round t |
| *Output*: $Q_t$ | // offspring population $Q_t$ |
| 1: **select** $sol_1$, $sol_2$ **from** $P_t$; | //two solutions selected from $P_t$ |
| 2: **crossover** ($sol_1$, $sol_2$); | |
| 3: **mutation** ( $offspring_1$ ); | |
| 4: **mutation** ( $offspring_2$ ); | |
| 5: **return** $Q_t$ | |

### 3.3.5. Main Phase of VNF mapping algorithm

The main phase of the proposed algorithm is to input the initial population $p_0$ and generate the offspring population $Q_0$. For this end, at first, a random initial population $p_0$ with size N is created and fed to the VNF mapping algorithm. Then, the sorting algorithm fast_non_dominated_sort is called to obtain the Pareto non-dominated leading set F=$\{F_1, F_2, …\}$. In order to generate the generation of offspring $Q_0$ with size N, the operators of binary tournament, crossover, mutation and the NSGA-II algorithm (shown in Algorithm 4) are used. The NSGA-II algorithm first combines the population $p_t$ and $Q_t$ to form a new population $R_t$ of size N. Then, the population $R_t$ is sorted using the sorting algorithm fast_non_dominated_sort. The main point in achieving elitism in NSGA-II is that it should search for solutions that belong to the best non-dominated set $F_1$. If there are fewer than N chromosomes in $F_1$, NSGA-II will definitely select all chromosomes of the set $F_1$ for the new population $p_{t+1}$. The remaining $p_{t+1}$ members are selected from $F_2$ solutions, followed by $F_3$, and so on until the last set is reached. To find the best solutions to fill all gaps in the population, the NSGA-II algorithm uses the sort () subroutine with the $\prec_n$ operator. Then, the make_new_pop () subroutine to make the selection, crossing and mutation and creating the next generation $Q_{t+1}$ is fed by the population $p_{t+1}$ with size N. Finally, the NSGA-II algorithm is called recursively until the required number of iterations is met. At this stage, the Pareto optimal front contains the desired solutions for the VNF mapping problem.

**Algorithm 4: The NSGA-II algorithm**

| | | |
|---|---|---|
| **Input: $P_0$** | | // initial population |
| 1: | $t = 0$ | |
| 2: | **while** (number of iterations not reached) $F = \textbf{\textit{fast\_non\_dominated\_sort}} (\textbf{\textit{p}}_t)$ | |
| 3: | $Q_t = \textbf{\textit{make\_new\_pop}} (\textbf{\textit{P}}_t)$ | /* use selection, crossover and mutation to create a new population $Q_t$ */ |
| 4: | $R_t = P_t \cup Q_t$ | |
| 5: | $F = \textbf{\textit{fast\_non\_dominated\_sort}} (\textbf{\textit{R}}_t)$ | // $F = (F_1, F_2, …)$, all non-dominated fronts of $R_t$ |
| 6: | $P_{t+1} = \emptyset$ | |
| 7: | $i=1$ | |
| 8: | **until** $|P_{t+1}| + |F_i| \leq N$ **do** | // until the parent population is filled |
| 9 | **crowding_distance_assignment** $(\textbf{\textit{F}}_i)$ | //calculate crowding distance in $F_i$ |
| 10 | $P_{t+1} = P_{t+1} \cup F_i$ | // include i th non-dominated front in the parent pop |
| 11 | $i = i + 1$ | // check the next front for inclusion |
| 12: | **sort** $(\textbf{\textit{F}}_i, \prec_n)$ | // sort in descending order using $\prec_n$ |

13: $\quad P_{t+1} = P_{t+1} \cup F_i[1:(N - |P_{t+1}|)]$  //choose the first $(N - |P_{t+1}|)]$ elements of $F_i$

14: $\quad Q_{t+1} = make\_new\_pop\ (P_{t+1})$

15: $\quad$ NSGA_II$(P_{t+1}, Q_{t+1})$

## 5. Simulation results

In this section, we evaluate the effectiveness of the proposed method (MODMVNF) in generating optimal possible solutions for the mapping problem of VNFs on physical nodes. In order to show the efficiency and performance of the our proposed method, we compare it with the particle swarm optimization algorithm. Simulation of the MODMVNF algorithm is run in MATLAB 2015b software and in a 64-bit system with X64-based processor, with Windows 10 Pro version, Intel Core i5-5500 u 2.4 GHz processor, 12 GB memory (RAM), NVIDIA GeForce 820M-2G graphics processor. Also, the graphic memory is 2 GB HDD.

To evaluate the performance of the MODMVNF algorithm, first, we consider a small physical network including 10 physical nodes, each of which has 16 TB of secondary memory and 1 GB of internal memory, and runs on a quad-core processor with a total processing capacity of 1000 MIPS and compare the results of our algorithm with integer linear programming (ILP) and PSO. Then, we will expand the number of physical hosts up to 30. As mentioned, at first, the physical structure of the problem is a non-path graph that contains 10 nodes as physical nodes that are connected by links (L). We assume that there is not necessarily a connecting link between each node, but there is at least one path between all nodes. In addition, $C_{cost_{u_j}}$ ، $M_{cost_{u_j}}$ و $S_{cost_{u_j}}$ for all 10 physical nodes are the same and equal to 3, 2 and 1, respectively. Also, we assume that each NF can be of a different type:

$\forall i \in N_{dc}:\quad i \in \{VM, process, I/O, \text{hardware}\}.$

The processing rate of each VM randomly from the MIPS set {250,500,750,750,1000,1000}, the processing rate of each process from the MIPS set {250,500,750,750,1000,1000,1000}, the processing rate of each I/O from the MIPS set {250,500,750,1000} and the processing rate of each hardware are selected from the set of MIPS {250,250,250,500,500,750,1000}. The required memory of each VM from the set of MB {128,256,256,512,512,512}, the memory of each process from the set of MB {128,256,512}, the memory of each I/O from the set of {128,256,512} MB and the memory of each hardware from the set of {128,256,512} MB, as well as the storage amount of each VM from the set { 256,512,1024,1024} MB, the storage amount of each process from the set{ 256,256,256,512,512,1024}MB, the storage amount of each I/O from the set { 256,512,1024} MB and the storage rate of each hardware from the set { 256,512,512,1024,1024,1024} MB is selected. The scenario used to calculate the number of physical nodes used and calculate the delay between these nodes as well as calculate the cost of VNFs mapping on the physical infrastructure includes several SFCs with different decompositions. We assume that each SFC has a maximum of 4 different partitions, which consist of a maximum of ten VNFs with different virtualization technologies. The initial population that we consider in our proposed algorithm includes N=60 solutions, whose initial mapping on physical nodes is done randomly. Our goal is to simultaneously map VNFs on physical nodes and minimize the cost of mapping, to obtain the shortest path between the used nodes (delay (latency) between nodes) by Dijkstra's algorithm.

Figure 6 shows the average mapping cost for the different SFCs. It is worth mentioning that while solving the problem of mapping VNFs and calculating the latency between used nodes, the best type of decomposition of each SG is also determined. In fact, during the execution of the mapping algorithm, the one with the lowest mapping cost is selected. To compare the MODMVNF algorithm with the PSO algorithm and ILP, we consider the results of the MODMVNF and PSO algorithms for each SFC as the average of members of the Pareto front $F_1$ and compare these results with the unique optimal solution of ILP. In these algorithms for each SFC, the average cost obtained from the mapping of VNFs by the members of the Pareto front $F_1$ is calculated and a number is obtained for each SFC in each algorithm. Figure 6 shows that the performance of the MODMVNF algorithm is close to that of the ILP approach, but it has superiority and better performance compared to the PSO algorithm, especially for more VNFs.

Figure 7 shows the latency among used nodes for algorithms MODMVNF, PSO, and ILP with different SFCs. In MODMVNF and PSO algorithms, for each SFC, the average latency between the physical nodes used by the members of the Pareto front $F_1$ is calculated and a number is obtained for each SFC, which we compare the obtained numbers by the two algorithms for each SFC with the obtained optimal solution by ILP. As seen in Figure 7, the MODMVNF algorithm has a lower time latency than the PSO algorithm and ILP. Therefore, it has a better performance. Figures 8 and 9 are the plots of the cumulative distribution of the MODMVNF, PSO algorithms simulation, and also ILP. These figures show the cumulative distribution of mapping cost VNFs on the used physical nodes and the cumulative distribution of latency between these nodes.

It is important to mention that the MODMVNF proposed algorithm simultaneously optimizes two objective functions $f_1$ and $f_2$ in equations 1 and 2, respectively. Probably if only $f_1$ or only $f_2$ is considered as the objective, better results will be obtained for the latency or the cost of mapping VNFs on physical nodes.

To illustrate the ability and efficiency of the MODMVNF algorithm for large networks, we perform simulations of the proposed algorithm for a network with 30 physical hosts and 10 SFCs. Figures 10-13 show the obtained results. According to these figures, we can see that the MODMVNF algorithm is efficient in large networks too. Also, the results show the MODMVNF algorithm is better than the PSO algorithm.

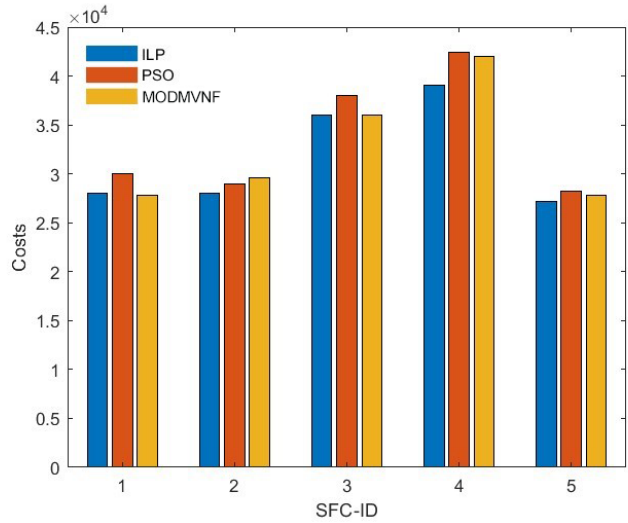

**Figure 6: The bar graph resulting from the MODMVNF, PSO algorithms, and ILP for the mapping cost of VNFs for 12 physical hosts**

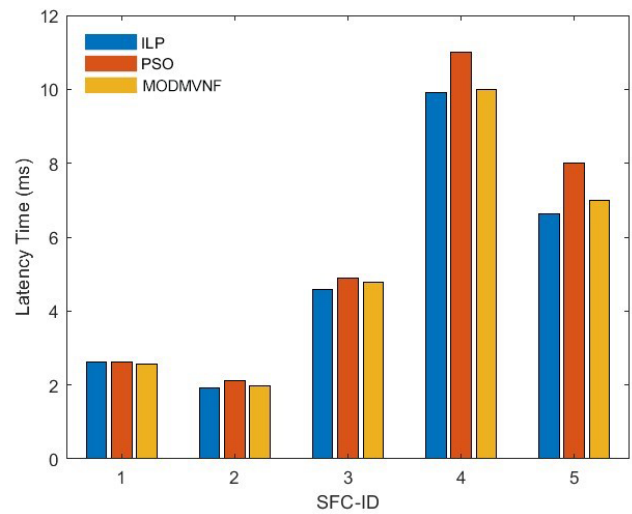

**Figure 7: The bar graph resulting from the MODMVNF, PSO algorithms, and ILP for the latency between used nodes for 12 physical hosts**

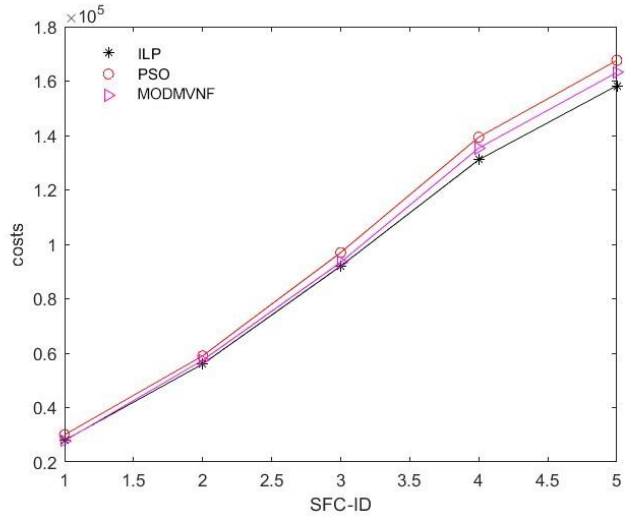

**Figure 8: The cumulative distribution from the MODMVNF, PSO algorithms and ILP for the mapping cost of VNFs for a network with 12 physical hosts**

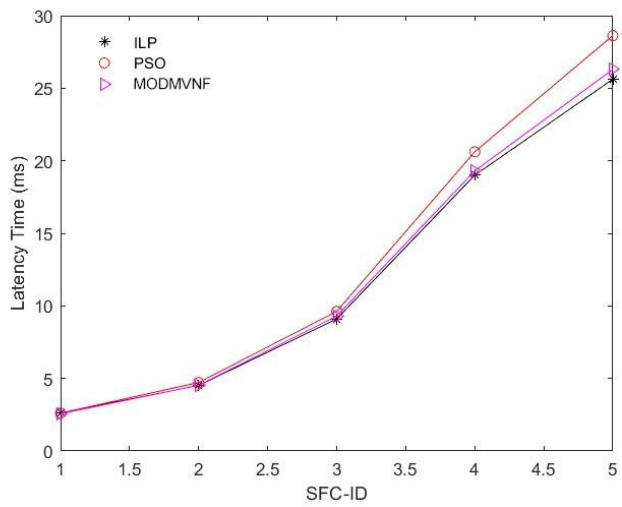

**Figure 9: The cumulative distribution from the MODMVNF, PSO algorithms and ILP for the latency between used nodes for a network with 12 physical hosts**

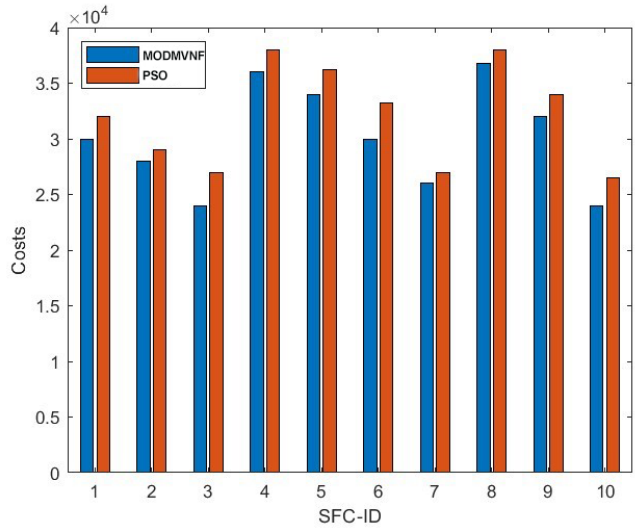

**Figure 10: The bar graph resulting from the MODMVNF algorithm and the PSO algorithm for the mapping cost of VNFs for a large network with 30 physical hosts**

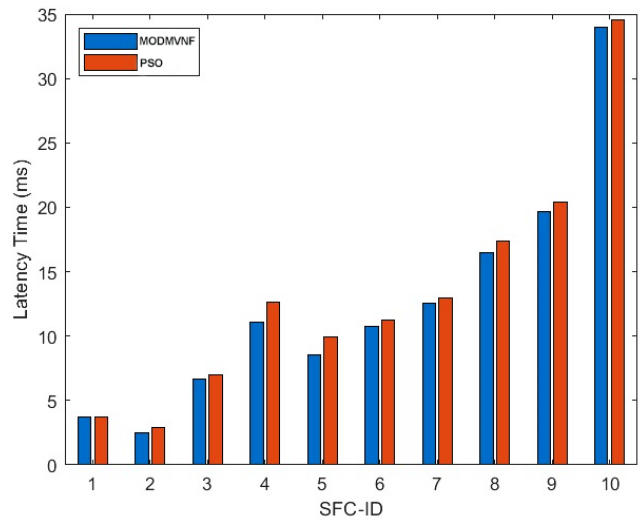

**Figure 11: The bar graph resulting from the MODMVNF algorithm and the PSO algorithm for the latency between used nodes for a large network with 30 physical hosts**

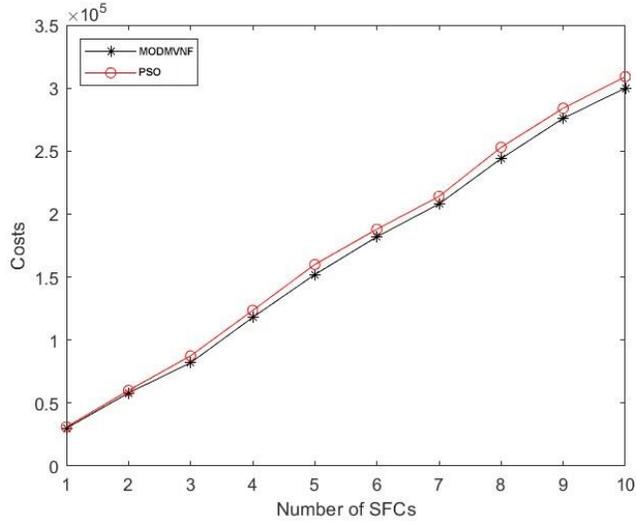

**Figure 12: The cumulative distribution from the MODMVNF algorithm and the PSO algorithm for the mapping cost of VNFs for a large network with 30 physical hosts**

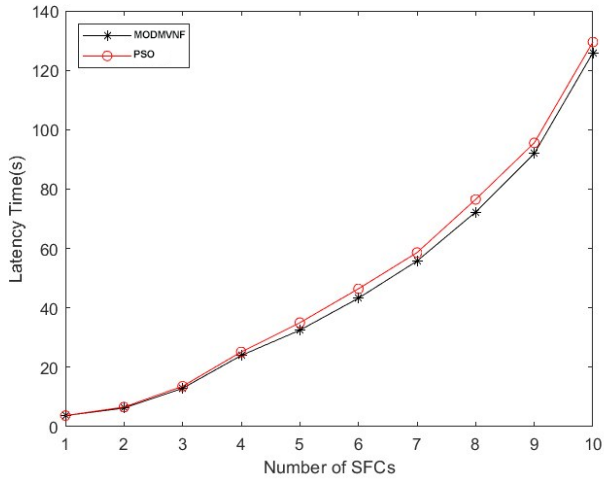

**Figure 13: The cumulative distribution from the MODMVNF algorithm and the PSO algorithm for the latency between used nodes for a large network with 30 physical hosts**

## 6. Conclusion

In this paper, we have presented a meta-heuristic solution based on NSGA-II to solve the mapping problem of VNFs in order to minimize the delay between the used physical nodes and the mapping cost. We assumed that the incoming requests are in the form of SFCs, each of which has ten VNFs, and each SFC has several decompositions, during the algorithm execution process, the best decomposition is selected so that the cost of mapping and the delay between physical nodes are minimized. The physical structure of the problem was considered as a non-path graph that included N nodes as physical nodes that are connected by links (L). We also assumed that there is not necessarily a link between two physical nodes, but there must be a path between two physical nodes. In addition, we assumed that the edges of the graph have a very large bandwidth capacity and do not impose any constraints on the VNF-P problem. The simulation results showed the efficiency and accuracy of the MODMVNF approach in terms of important criteria such as average delay, cost of mapping VNFs on physical infrastructure, and choosing the best decomposition compared to PSO.